\documentclass[prl,twocolumn, aps,superscriptaddress,notitlepage]{revtex4-2}
\usepackage{amssymb,amsfonts,amsmath,amstext,graphicx,hyperref,lipsum,empheq}
\usepackage{float}
\usepackage[normalem]{ulem}
\hypersetup{
	colorlinks=true,
	linkcolor=blue,
	filecolor=magenta,      
	urlcolor=cyan,
}
\usepackage{amsthm}
\usepackage{outlines}

\usepackage[dvipsnames]{xcolor}

\newcommand{\beq}[0]{\begin{equation}}
\newcommand{\eeq}[0]{\end{equation}}

\def\be{\begin{equation}}
\def\ee{\end{equation}}
\def\bea{\begin{eqnarray}}
\def\eea{\end{eqnarray}}
\newcommand{\ba}{\begin{eqnarray}}
\newcommand{\ea}{\end{eqnarray}}
\usepackage{hyperref}

\def\BraVert{\egroup\,\mid\,\bgroup}

\definecolor{myblue}{rgb}{.8, .8, 1}

\usepackage{lineno}

\begin{document}

\title{The promises and challenges of many-body quantum technologies: a focus on quantum engines }

\author{Victor Mukherjee}\email{mukherjeev@iiserbpr.ac.in}
\affiliation{Department of Physical Sciences, Indian Institute of Science Education and Research Berhampur, Berhampur, 760010, India}

\author{Uma Divakaran}\email{uma@iitpkd.ac.in}
\affiliation{Department of Physics, Indian Institute of Technology Palakkad, Palakkad 678623, India}

\begin{abstract}
\section*{Abstract}
Can many-body systems be beneficial to designing quantum technologies? We address this question by examining quantum engines, where recent studies indicate potential benefits through the harnessing of many-body effects, such as divergences close to phase transitions. However, open questions remain regarding their real-world applications.

\end{abstract}

\maketitle

\date{\today}

\date{\today}

\section*{Introduction}

Quantum technologies (QTs) attempt to harness quantum systems to go beyond the bounds set by conventional classical technologies. Quantum many-body systems exhibit emergent properties of many interacting quantum particles, often lacking classical analogs. It is thus a relevant question to ask if and how quantum many-body properties might be useful for the development of QTs. In particular, recent studies have shown that the development of quantum engines might benefit from many-body physics.

Yet, analysis and control of many-body quantum systems can be highly non-trivial, owing to
the large dimension of the associated Hilbert space, which diverges exponentially with the system size. 
Nevertheless, inter-particle interactions and large system-sizes in many-body quantum systems may offer several possibilities  that may not exist in their few-body counterparts. This provides researchers with more avenues for designing novel and high-performing QTs \cite{mukherjee21many}. Consequently, combining the fields of many-body physics and QTs to develop many-body QTs, can disrupt the technologies that drive the modern world.
For example,  enhancement in quantum Fisher information close to a  continuous quantum phase transition (PT), which occurs only in many-body quantum systems, has been shown to be extremely beneficial for engineering highly accurate quantum magnetometers \cite{montenegro21global} and quantum thermometers \cite{zhang22non}. Similar enhancements due to many-body cooperative dynamics may also allow us to perform high-precision quantum
thermometry \cite{latune20collective}. 
The entanglement properties of many-body localized systems may be used to design quantum batteries with high {\it ergotropy} or work capacity \cite{rossini19many}, while the energy gap statistics of such systems may enhance the reliability of  quantum heat engines \cite{halpern19quantum}.
Furthermore, the intriguing possibility remains that other many-body effects, which have hitherto not been explored extensively in the context of QTs, such as quantum scars \cite{zhang23many}, 
can present new possibilities for developing high-performing QTs. Recent advances in machine learning can also play a major role in optimally controlling many-body QTs  \cite{erdman22identifying}.   

Several platforms have been shown to be suitable for experimentally realizing many of the many-body QTs discussed above. For example, quantum simulators modeled by superconducting qubits \cite{king22coherent} and Rydberg atoms \cite{ebadi21quantum} have been used for simulating many-body spin systems driven through 
quantum PTs.
As shown in a recent experiment using superfluid Helium, connecting two time-crystals can enable us to make qubits, which can be the building blocks of quantum computers and quantum information processing setups \cite{autti21AC}. 
Experimentally, time-crystals can  be realized in optical setups at room temperatures as well \cite{taheri22all}; this may pave the way for easier production and real-world applications of quantum time-crystals, such as for designing quantum engines \cite{carollo20nonequilibrium}. Recently, the quantum statistics of ultracold $^{6}$Li atoms have been used to experimentally realize a many-body quantum engine; in contrast to conventional heat engines powered by thermal baths, here energy stemming from the Pauli exclusion principle was cyclically converted into output work \cite{koch23a}.

Below, we focus on the technical aspects of a widely-studied QT, namely quantum engines, to have a deeper understanding of the role played by many-body physics in this field. \\

\section*{Many-body effects in quantum engines}

\begin{figure}[h]
         \centering
         \includegraphics[width=0.9\linewidth]{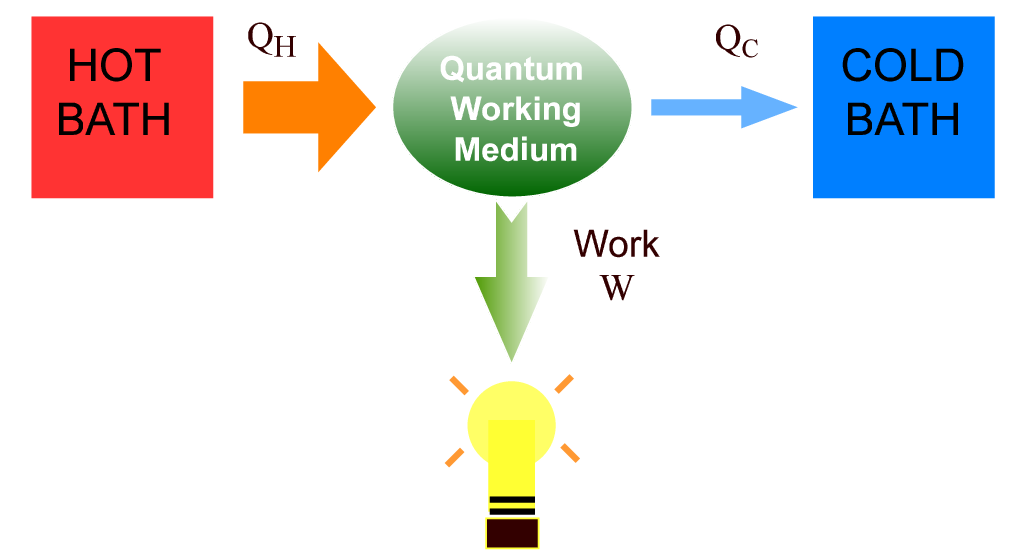}
         \caption{{\bf A quantum heat engine:} Schematic diagram of a quantum heat engine. Heat $Q_h$ flows from the hot bath to the quantum working medium; a part $Q_c$ of the heat flows from the working medium to the cold bath. The rest of the energy is converted to output work $W$, which can be used to do some useful task, such as light a bulb.}
        \label{fig:QHE}
\end{figure}

The interplay between unitary and non-unitary dynamics can make quantum engines an ideal platform for studying the thermodynamics of QTs.  
In general, a quantum heat engine (QHE) is modeled by a periodically modulated quantum system, termed the working medium (WM), coupled to a hot and a cold bath. Heat flows from the hot bath to the WM; a part of the heat flows to the cold bath, while the rest of the heat is converted to output work (see Fig. \ref{fig:QHE}). 
The amount of work output per unit time is known as the power, while the ratio of the output work to the input heat is the efficiency. 
One of the primary aims of research in this field is to enhance the output work, power, and efficiency of a QHE, while reducing the fluctuations in the output of the QHE.  Experimentally, QHEs have been realized using different systems, including nitrogen-vacancy centers in diamonds \cite{klatzow19experimental} and nuclear magnetic resonance setups \cite{peterson19experimental}.  As we discuss below, several studies have shown that many-body WMs may indeed fulfill  many of the ambitious aims described above \cite {mukherjee21many}.

{\bf Cooperative large-spin effects:} Cooperative large-spin effects (CLSEs) can arise when multiple identical spins interact with the same environment \cite{ niedenzu18cooperative, latune20collective, mukherjee21many, jaseem23quadratic}, which, for example, can be realized using atoms in an optical cavity \cite{norcia18cavity}. The  spins can then share excitations among themselves, and under suitable conditions act as a single large spin, which may result in the setup having a high specific heat \cite{latune20collective}.
Recently, researchers have  proposed to harness this enhancement in specific heat to significantly increase the work, power output \cite{niedenzu18cooperative, latune20collective} and the reliability \cite{jaseem23quadratic} in  QHEs based on multiple identical spins collectively coupled to thermal baths.

One major advantage of using CLSEs to analyze many-body QHEs is the simplicity of their analytical treatment in terms of the dynamics of the effective single large spin introduced above. However, in spite of the analytical simplicity and the possibility of realizing them experimentally using well-established quantum-optical platforms \cite{norcia18cavity},   CLSEs may not be able to provide significant insights into QHEs with more generic setups, which are not describable by  single large spins. For that, we turn to the next example.

{\bf Phase transitions:} Phase transitions in many-body systems are associated with non-analyticities, which give rise to several highly interesting effects, such as divergences in different quantities.   Consequently, the potential of PTs to aid in the operation of QHEs has also been explored in several recent works \cite{mukherjee21many}. For example, in general, a heat engine operating with an efficiency close to the Carnot limit delivers vanishingly small power output. However,  in Ref. \cite{campisi16the}, the authors showed that diverging specific heat close to a continuous PT may enable us to circumvent this significant limitation. The authors considered an Otto engine with the hot bath at a critical temperature, such that the many-body WM remains on the verge of a continuous PT  when coupled to this bath. The authors 
showed that
choosing a WM material with suitable critical exponents, such as DY$_2$Ti$_2$O$_7$, may allow us to model a many-body heat engine that delivers non-zero power output, even close to the Carnot limit of maximum efficiency. 

In spite of the remarkable results reported in \cite{campisi16the}, the power of a critical QHE still decreases as we approach the Carnot limit, finally vanishing at the Carnot limit. Furthermore, non-adiabatic excitations in engines operating close to quantum PTs can be detrimental to their operation \cite{mukherjee21many}, and  large fluctuations in the output of such engines can significantly limit their viability for industrial applications \cite{holubec17work}. 
Consequently, the relevance of critical QHEs for real-world usage is still unclear, and more rigorous research is needed to understand if and how PTs can indeed be beneficial for developing QHEs.

{\bf Time-crystals:} 
A recently discovered dynamical many-body state that can be highly relevant to designing QTs is a time crystal. Time crystals are categorized into discrete and continuous depending on whether they spontaneously break discrete or continuous time-translational symmetry. They are characterized by a persistent oscillatory behavior. Discrete time crystals have been realized in both closed and open systems under an external periodic drive. 
The so-called dissipative discrete time crystals realized in open systems can be viewed as a form of quantum engine, where the energy input from a periodic  drive is partially converted to output work, while  delivering heat to a cold bath  \cite{zaletel23colloquium}.
 Continuous time crystals (CTC) do not require a periodic drive and exhibit spontaneous time-periodic behavior. A CTC incorporates autonomous operation naturally and therefore can play an important role in the development of many-body autonomous quantum engines (AQEs). Such many-body AQEs can be an integral part of different near and long-term applications,  such as  for powering microscopic robots for drug delivery inside human bodies.

\begin{figure}[h]
         \centering
         \includegraphics[width=0.9\linewidth]{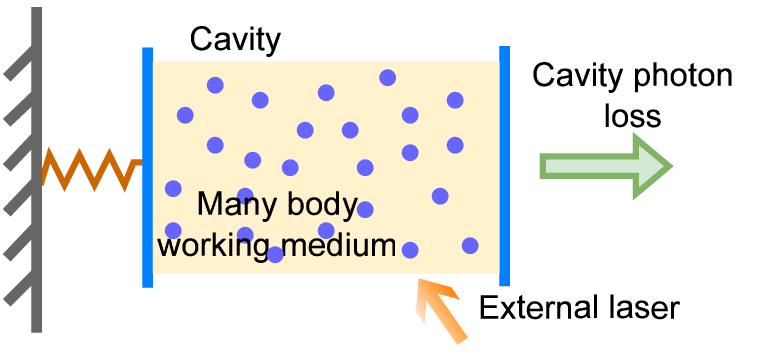}
         \caption{{\bf A time-crystal quantum engine:} Schematic diagram of a time-crystal quantum engine proposed in \cite{carollo20nonequilibrium}. An external laser shines on two-level atoms (in blue) inside a cavity. One wall of the cavity is a movable mirror attached to a microspring (on the left).  The mirror can oscillate around its equilibrium position due to thermal fluctuations and radiation pressure inside the cavity, thereby resulting in a work output.}
        \label{fig:TC}
\end{figure}

 Indeed, in a recent work, researchers have proposed the idea of using the periodic behavior of a CTC to model a many-body AQE, comprising atoms illuminated by light inside an optical cavity, and exchanging energy with a small movable mirror  (see Fig. \ref{fig:TC}). Under suitable conditions, a CTC may form, such that the mirror shows periodic oscillations even without any external driving, thus resulting in an engine with work output \cite{carollo20nonequilibrium}. 

As seen above, the properties of many-body WMs that facilitate the modeling of quantum engines are varied, with their own advantages and disadvantages. Specially, the role played by time-crystals is still less explored and may provide an exciting field of research in the near future. 
The effects of integrability,  which allows studying multiple interacting particles in terms of non-interacting quasi-particles
\cite{revathy22bath}, and topological properties,  which are robust to local changes in a many-body WM \cite{yunt20internal}, on the operation of QHEs are some of the other less studied topics that may lead to interesting areas of research.

\section{The way ahead}

In the short term, further theoretical research on possible many-body effects that can be used for designing QTs, such as integrability \cite{revathy22bath},  and the relevant costs involved,  for example the energetic costs for design and control of such many-body QTs \cite{campbell17trade, mukherjee21many}, can give us a deeper understanding of the potential advantages of many-body QTs over their classical counterparts. Relatedly, control of many-body quantum systems in the presence of dissipation can be one of the main challenges behind experimental realization of many-body QTs. Therefore concurrent
research on control of many-body open quantum systems can be essential for the experimental development of many-body QTs \cite{bai20floquet, erdman22identifying}.

In the long term, practical large-scale application of many-body QTs would depend on the commercial production of such technologies. Therefore rigorous studies
on quantum materials which are easily available, controllable, and robust to dissipative effects of the environment, are crucial for  the eventual commercial production of many-body QTs. For instance, many-body materials which can store significant amount of ergotropy, i.e., energy which can be converted to work, 
 and are robust to energy leakage through external dissipation, may be ideal for the development of quantum batteries \cite{rossini19many} and therefore can be a significant practical step towards the  transition to renewable energy sources. 
 
 To conclude, the field of many-body QTs promises many exciting developments in the near future, and more rigorous research is needed to understand the full potential of this field.

 \section*{Acknowledgement}
 V.M. acknowledges support from Science and Engineering Research Board (SERB) through MATRICS (Project No. MTR/2021/000055) and a Seed Grant from IISER
Berhampur. 

\section*{Author contributions}
V.M. and U.D. discussed and wrote the comment.

\section*{Competing interests}
The authors declare no competing interests.

\section*{Additional information}
{\bf Correspondence} and requests for materials should be addressed to Victor Mukherjee

\color{RoyalBlue}
\providecommand{\noopsort}[1]{}\providecommand{\singleletter}[1]{#1}%


\begin{thebibliography}{26}%
\makeatletter
\providecommand \@ifxundefined [1]{%
 \@ifx{#1\undefined}
}%
\providecommand \@ifnum [1]{%
 \ifnum #1\expandafter \@firstoftwo
 \else \expandafter \@secondoftwo
 \fi
}%
\providecommand \@ifx [1]{%
 \ifx #1\expandafter \@firstoftwo
 \else \expandafter \@secondoftwo
 \fi
}%
\providecommand \natexlab [1]{#1}%
\providecommand \enquote  [1]{``#1''}%
\providecommand \bibnamefont  [1]{#1}%
\providecommand \bibfnamefont [1]{#1}%
\providecommand \citenamefont [1]{#1}%
\providecommand \href@noop [0]{\@secondoftwo}%
\providecommand \href [0]{\begingroup \@sanitize@url \@href}%
\providecommand \@href[1]{\@@startlink{#1}\@@href}%
\providecommand \@@href[1]{\endgroup#1\@@endlink}%
\providecommand \@sanitize@url [0]{\catcode `\\12\catcode `\$12\catcode
  `\&12\catcode `\#12\catcode `\^12\catcode `\_12\catcode `\%12\relax}%
\providecommand \@@startlink[1]{}%
\providecommand \@@endlink[0]{}%
\providecommand \url  [0]{\begingroup\@sanitize@url \@url }%
\providecommand \@url [1]{\endgroup\@href {#1}{\urlprefix }}%
\providecommand \urlprefix  [0]{URL }%
\providecommand \Eprint [0]{\href }%
\providecommand \doibase [0]{https://doi.org/}%
\providecommand \selectlanguage [0]{\@gobble}%
\providecommand \bibinfo  [0]{\@secondoftwo}%
\providecommand \bibfield  [0]{\@secondoftwo}%
\providecommand \translation [1]{[#1]}%
\providecommand \BibitemOpen [0]{}%
\providecommand \bibitemStop [0]{}%
\providecommand \bibitemNoStop [0]{.\EOS\space}%
\providecommand \EOS [0]{\spacefactor3000\relax}%
\providecommand \BibitemShut  [1]{\csname bibitem#1\endcsname}%
\let\auto@bib@innerbib\@empty
\bibitem [{\citenamefont {Mukherjee}\ and\ \citenamefont
  {Divakaran}(2021)}]{mukherjee21many}%
  \BibitemOpen
  \bibfield  {author} {\bibinfo {author} {\bibfnamefont {V.}~\bibnamefont
  {Mukherjee}}\ and\ \bibinfo {author} {\bibfnamefont {U.}~\bibnamefont
  {Divakaran}},\ }\bibfield  {title} {\bibinfo {title} {Many-body quantum
  thermal machines},\ }\href {https://doi.org/10.1088/1361-648x/ac1b60}
  {\bibfield  {journal} {\bibinfo  {journal} {Journal of Physics: Condensed
  Matter}\ }\textbf {\bibinfo {volume} {33}},\ \bibinfo {pages} {454001}
  (\bibinfo {year} {2021})}\BibitemShut {NoStop}%
\bibitem [{\citenamefont {Montenegro}\ \emph {et~al.}(2021)\citenamefont
  {Montenegro}, \citenamefont {Mishra},\ and\ \citenamefont
  {Bayat}}]{montenegro21global}%
  \BibitemOpen
  \bibfield  {author} {\bibinfo {author} {\bibfnamefont {V.}~\bibnamefont
  {Montenegro}}, \bibinfo {author} {\bibfnamefont {U.}~\bibnamefont {Mishra}},\
  and\ \bibinfo {author} {\bibfnamefont {A.}~\bibnamefont {Bayat}},\ }\bibfield
   {title} {\bibinfo {title} {Global sensing and its impact for quantum
  many-body probes with criticality},\ }\href
  {https://doi.org/10.1103/PhysRevLett.126.200501} {\bibfield  {journal}
  {\bibinfo  {journal} {Phys. Rev. Lett.}\ }\textbf {\bibinfo {volume} {126}},\
  \bibinfo {pages} {200501} (\bibinfo {year} {2021})}\BibitemShut {NoStop}%
\bibitem [{\citenamefont {Zhang}\ \emph {et~al.}(2022)\citenamefont {Zhang},
  \citenamefont {Chen}, \citenamefont {Bai}, \citenamefont {Wu},\ and\
  \citenamefont {An}}]{zhang22non}%
  \BibitemOpen
  \bibfield  {author} {\bibinfo {author} {\bibfnamefont {N.}~\bibnamefont
  {Zhang}}, \bibinfo {author} {\bibfnamefont {C.}~\bibnamefont {Chen}},
  \bibinfo {author} {\bibfnamefont {S.-Y.}\ \bibnamefont {Bai}}, \bibinfo
  {author} {\bibfnamefont {W.}~\bibnamefont {Wu}},\ and\ \bibinfo {author}
  {\bibfnamefont {J.-H.}\ \bibnamefont {An}},\ }\bibfield  {title} {\bibinfo
  {title} {Non-markovian quantum thermometry},\ }\href
  {https://doi.org/10.1103/PhysRevApplied.17.034073} {\bibfield  {journal}
  {\bibinfo  {journal} {Phys. Rev. Appl.}\ }\textbf {\bibinfo {volume} {17}},\
  \bibinfo {pages} {034073} (\bibinfo {year} {2022})}\BibitemShut {NoStop}%
\bibitem [{\citenamefont {Latune}\ \emph {et~al.}(2020)\citenamefont {Latune},
  \citenamefont {Sinayskiy},\ and\ \citenamefont
  {Petruccione}}]{latune20collective}%
  \BibitemOpen
  \bibfield  {author} {\bibinfo {author} {\bibfnamefont {C.~L.}\ \bibnamefont
  {Latune}}, \bibinfo {author} {\bibfnamefont {I.}~\bibnamefont {Sinayskiy}},\
  and\ \bibinfo {author} {\bibfnamefont {F.}~\bibnamefont {Petruccione}},\
  }\bibfield  {title} {\bibinfo {title} {Collective heat capacity for quantum
  thermometry and quantum engine enhancements},\ }\href
  {https://doi.org/10.1088/1367-2630/aba463} {\bibfield  {journal} {\bibinfo
  {journal} {New Journal of Physics}\ }\textbf {\bibinfo {volume} {22}},\
  \bibinfo {pages} {083049} (\bibinfo {year} {2020})}\BibitemShut {NoStop}%
\bibitem [{\citenamefont {Rossini}\ \emph {et~al.}(2019)\citenamefont
  {Rossini}, \citenamefont {Andolina},\ and\ \citenamefont
  {Polini}}]{rossini19many}%
  \BibitemOpen
  \bibfield  {author} {\bibinfo {author} {\bibfnamefont {D.}~\bibnamefont
  {Rossini}}, \bibinfo {author} {\bibfnamefont {G.~M.}\ \bibnamefont
  {Andolina}},\ and\ \bibinfo {author} {\bibfnamefont {M.}~\bibnamefont
  {Polini}},\ }\bibfield  {title} {\bibinfo {title} {Many-body localized
  quantum batteries},\ }\href {https://doi.org/10.1103/PhysRevB.100.115142}
  {\bibfield  {journal} {\bibinfo  {journal} {Phys. Rev. B}\ }\textbf {\bibinfo
  {volume} {100}},\ \bibinfo {pages} {115142} (\bibinfo {year}
  {2019})}\BibitemShut {NoStop}%
\bibitem [{\citenamefont {Yunger~Halpern}\ \emph {et~al.}(2019)\citenamefont
  {Yunger~Halpern}, \citenamefont {White}, \citenamefont {Gopalakrishnan},\
  and\ \citenamefont {Refael}}]{halpern19quantum}%
  \BibitemOpen
  \bibfield  {author} {\bibinfo {author} {\bibfnamefont {N.}~\bibnamefont
  {Yunger~Halpern}}, \bibinfo {author} {\bibfnamefont {C.~D.}\ \bibnamefont
  {White}}, \bibinfo {author} {\bibfnamefont {S.}~\bibnamefont
  {Gopalakrishnan}},\ and\ \bibinfo {author} {\bibfnamefont {G.}~\bibnamefont
  {Refael}},\ }\bibfield  {title} {\bibinfo {title} {Quantum engine based on
  many-body localization},\ }\href {https://doi.org/10.1103/PhysRevB.99.024203}
  {\bibfield  {journal} {\bibinfo  {journal} {Phys. Rev. B}\ }\textbf {\bibinfo
  {volume} {99}},\ \bibinfo {pages} {024203} (\bibinfo {year}
  {2019})}\BibitemShut {NoStop}%
\bibitem [{\citenamefont {Zhang}\ \emph {et~al.}(2023)\citenamefont {Zhang},
  \citenamefont {Dong}, \citenamefont {Gao}, \citenamefont {Zhao},
  \citenamefont {Hao}, \citenamefont {Desaules}, \citenamefont {Guo},
  \citenamefont {Chen}, \citenamefont {Deng}, \citenamefont {Liu},
  \citenamefont {Ren}, \citenamefont {Yao}, \citenamefont {Zhang},
  \citenamefont {Xu}, \citenamefont {Wang}, \citenamefont {Jin}, \citenamefont
  {Zhu}, \citenamefont {Zhang}, \citenamefont {Li}, \citenamefont {Song},
  \citenamefont {Wang}, \citenamefont {Liu}, \citenamefont {Papi{\'{c}}},
  \citenamefont {Ying}, \citenamefont {Wang},\ and\ \citenamefont
  {Lai}}]{zhang23many}%
  \BibitemOpen
  \bibfield  {author} {\bibinfo {author} {\bibfnamefont {P.}~\bibnamefont
  {Zhang}}, \bibinfo {author} {\bibfnamefont {H.}~\bibnamefont {Dong}},
  \bibinfo {author} {\bibfnamefont {Y.}~\bibnamefont {Gao}}, \bibinfo {author}
  {\bibfnamefont {L.}~\bibnamefont {Zhao}}, \bibinfo {author} {\bibfnamefont
  {J.}~\bibnamefont {Hao}}, \bibinfo {author} {\bibfnamefont {J.-Y.}\
  \bibnamefont {Desaules}}, \bibinfo {author} {\bibfnamefont {Q.}~\bibnamefont
  {Guo}}, \bibinfo {author} {\bibfnamefont {J.}~\bibnamefont {Chen}}, \bibinfo
  {author} {\bibfnamefont {J.}~\bibnamefont {Deng}}, \bibinfo {author}
  {\bibfnamefont {B.}~\bibnamefont {Liu}}, \bibinfo {author} {\bibfnamefont
  {W.}~\bibnamefont {Ren}}, \bibinfo {author} {\bibfnamefont {Y.}~\bibnamefont
  {Yao}}, \bibinfo {author} {\bibfnamefont {X.}~\bibnamefont {Zhang}}, \bibinfo
  {author} {\bibfnamefont {S.}~\bibnamefont {Xu}}, \bibinfo {author}
  {\bibfnamefont {K.}~\bibnamefont {Wang}}, \bibinfo {author} {\bibfnamefont
  {F.}~\bibnamefont {Jin}}, \bibinfo {author} {\bibfnamefont {X.}~\bibnamefont
  {Zhu}}, \bibinfo {author} {\bibfnamefont {B.}~\bibnamefont {Zhang}}, \bibinfo
  {author} {\bibfnamefont {H.}~\bibnamefont {Li}}, \bibinfo {author}
  {\bibfnamefont {C.}~\bibnamefont {Song}}, \bibinfo {author} {\bibfnamefont
  {Z.}~\bibnamefont {Wang}}, \bibinfo {author} {\bibfnamefont {F.}~\bibnamefont
  {Liu}}, \bibinfo {author} {\bibfnamefont {Z.}~\bibnamefont {Papi{\'{c}}}},
  \bibinfo {author} {\bibfnamefont {L.}~\bibnamefont {Ying}}, \bibinfo {author}
  {\bibfnamefont {H.}~\bibnamefont {Wang}},\ and\ \bibinfo {author}
  {\bibfnamefont {Y.-C.}\ \bibnamefont {Lai}},\ }\bibfield  {title} {\bibinfo
  {title} {Many-body hilbert space scarring on a superconducting processor},\
  }\href {https://doi.org/10.1038/s41567-022-01784-9} {\bibfield  {journal}
  {\bibinfo  {journal} {Nature Physics}\ }\textbf {\bibinfo {volume} {19}},\
  \bibinfo {pages} {120} (\bibinfo {year} {2023})}\BibitemShut {NoStop}%
\bibitem [{\citenamefont {Erdman}\ and\ \citenamefont
  {No{\'e}}(2022)}]{erdman22identifying}%
  \BibitemOpen
  \bibfield  {author} {\bibinfo {author} {\bibfnamefont {P.~A.}\ \bibnamefont
  {Erdman}}\ and\ \bibinfo {author} {\bibfnamefont {F.}~\bibnamefont
  {No{\'e}}},\ }\bibfield  {title} {\bibinfo {title} {Identifying optimal
  cycles in quantum thermal machines with reinforcement-learning},\ }\href
  {https://doi.org/10.1038/s41534-021-00512-0} {\bibfield  {journal} {\bibinfo
  {journal} {npj Quantum Information}\ }\textbf {\bibinfo {volume} {8}},\
  \bibinfo {pages} {1} (\bibinfo {year} {2022})}\BibitemShut {NoStop}%
\bibitem [{\citenamefont {King}\ \emph {et~al.}(2022)\citenamefont {King},
  \citenamefont {Suzuki}, \citenamefont {Raymond}, \citenamefont {Zucca},
  \citenamefont {Lanting}, \citenamefont {Altomare}, \citenamefont {Berkley},
  \citenamefont {Ejtemaee}, \citenamefont {Hoskinson}, \citenamefont {Huang},
  \citenamefont {Ladizinsky}, \citenamefont {MacDonald}, \citenamefont
  {Marsden}, \citenamefont {Oh}, \citenamefont {Poulin-Lamarre}, \citenamefont
  {Reis}, \citenamefont {Rich}, \citenamefont {Sato}, \citenamefont
  {Whittaker}, \citenamefont {Yao}, \citenamefont {Harris}, \citenamefont
  {Lidar}, \citenamefont {Nishimori},\ and\ \citenamefont
  {Amin}}]{king22coherent}%
  \BibitemOpen
  \bibfield  {author} {\bibinfo {author} {\bibfnamefont {A.~D.}\ \bibnamefont
  {King}}, \bibinfo {author} {\bibfnamefont {S.}~\bibnamefont {Suzuki}},
  \bibinfo {author} {\bibfnamefont {J.}~\bibnamefont {Raymond}}, \bibinfo
  {author} {\bibfnamefont {A.}~\bibnamefont {Zucca}}, \bibinfo {author}
  {\bibfnamefont {T.}~\bibnamefont {Lanting}}, \bibinfo {author} {\bibfnamefont
  {F.}~\bibnamefont {Altomare}}, \bibinfo {author} {\bibfnamefont {A.~J.}\
  \bibnamefont {Berkley}}, \bibinfo {author} {\bibfnamefont {S.}~\bibnamefont
  {Ejtemaee}}, \bibinfo {author} {\bibfnamefont {E.}~\bibnamefont {Hoskinson}},
  \bibinfo {author} {\bibfnamefont {S.}~\bibnamefont {Huang}}, \bibinfo
  {author} {\bibfnamefont {E.}~\bibnamefont {Ladizinsky}}, \bibinfo {author}
  {\bibfnamefont {A.~J.~R.}\ \bibnamefont {MacDonald}}, \bibinfo {author}
  {\bibfnamefont {G.}~\bibnamefont {Marsden}}, \bibinfo {author} {\bibfnamefont
  {T.}~\bibnamefont {Oh}}, \bibinfo {author} {\bibfnamefont {G.}~\bibnamefont
  {Poulin-Lamarre}}, \bibinfo {author} {\bibfnamefont {M.}~\bibnamefont
  {Reis}}, \bibinfo {author} {\bibfnamefont {C.}~\bibnamefont {Rich}}, \bibinfo
  {author} {\bibfnamefont {Y.}~\bibnamefont {Sato}}, \bibinfo {author}
  {\bibfnamefont {J.~D.}\ \bibnamefont {Whittaker}}, \bibinfo {author}
  {\bibfnamefont {J.}~\bibnamefont {Yao}}, \bibinfo {author} {\bibfnamefont
  {R.}~\bibnamefont {Harris}}, \bibinfo {author} {\bibfnamefont {D.~A.}\
  \bibnamefont {Lidar}}, \bibinfo {author} {\bibfnamefont {H.}~\bibnamefont
  {Nishimori}},\ and\ \bibinfo {author} {\bibfnamefont {M.~H.}\ \bibnamefont
  {Amin}},\ }\bibfield  {title} {\bibinfo {title} {Coherent quantum annealing
  in a programmable 2,000{\thinspace}qubit ising chain},\ }\href
  {https://doi.org/10.1038/s41567-022-01741-6} {\bibfield  {journal} {\bibinfo
  {journal} {Nature Physics}\ }\textbf {\bibinfo {volume} {18}},\ \bibinfo
  {pages} {1324} (\bibinfo {year} {2022})}\BibitemShut {NoStop}%
\bibitem [{\citenamefont {Ebadi}\ \emph {et~al.}(2021)\citenamefont {Ebadi},
  \citenamefont {Wang}, \citenamefont {Levine}, \citenamefont {Keesling},
  \citenamefont {Semeghini}, \citenamefont {Omran}, \citenamefont {Bluvstein},
  \citenamefont {Samajdar}, \citenamefont {Pichler}, \citenamefont {Ho},
  \citenamefont {Choi}, \citenamefont {Sachdev}, \citenamefont {Greiner},
  \citenamefont {Vuleti{\'{c}}},\ and\ \citenamefont {Lukin}}]{ebadi21quantum}%
  \BibitemOpen
  \bibfield  {author} {\bibinfo {author} {\bibfnamefont {S.}~\bibnamefont
  {Ebadi}}, \bibinfo {author} {\bibfnamefont {T.~T.}\ \bibnamefont {Wang}},
  \bibinfo {author} {\bibfnamefont {H.}~\bibnamefont {Levine}}, \bibinfo
  {author} {\bibfnamefont {A.}~\bibnamefont {Keesling}}, \bibinfo {author}
  {\bibfnamefont {G.}~\bibnamefont {Semeghini}}, \bibinfo {author}
  {\bibfnamefont {A.}~\bibnamefont {Omran}}, \bibinfo {author} {\bibfnamefont
  {D.}~\bibnamefont {Bluvstein}}, \bibinfo {author} {\bibfnamefont
  {R.}~\bibnamefont {Samajdar}}, \bibinfo {author} {\bibfnamefont
  {H.}~\bibnamefont {Pichler}}, \bibinfo {author} {\bibfnamefont {W.~W.}\
  \bibnamefont {Ho}}, \bibinfo {author} {\bibfnamefont {S.}~\bibnamefont
  {Choi}}, \bibinfo {author} {\bibfnamefont {S.}~\bibnamefont {Sachdev}},
  \bibinfo {author} {\bibfnamefont {M.}~\bibnamefont {Greiner}}, \bibinfo
  {author} {\bibfnamefont {V.}~\bibnamefont {Vuleti{\'{c}}}},\ and\ \bibinfo
  {author} {\bibfnamefont {M.~D.}\ \bibnamefont {Lukin}},\ }\bibfield  {title}
  {\bibinfo {title} {Quantum phases of matter on a 256-atom programmable
  quantum simulator},\ }\href {https://doi.org/10.1038/s41586-021-03582-4}
  {\bibfield  {journal} {\bibinfo  {journal} {Nature}\ }\textbf {\bibinfo
  {volume} {595}},\ \bibinfo {pages} {227} (\bibinfo {year}
  {2021})}\BibitemShut {NoStop}%
\bibitem [{\citenamefont {Autti}\ \emph {et~al.}(2021)\citenamefont {Autti},
  \citenamefont {Heikkinen}, \citenamefont {M{\"a}kinen}, \citenamefont
  {Volovik}, \citenamefont {Zavjalov},\ and\ \citenamefont
  {Eltsov}}]{autti21AC}%
  \BibitemOpen
  \bibfield  {author} {\bibinfo {author} {\bibfnamefont {S.}~\bibnamefont
  {Autti}}, \bibinfo {author} {\bibfnamefont {P.~J.}\ \bibnamefont
  {Heikkinen}}, \bibinfo {author} {\bibfnamefont {J.~T.}\ \bibnamefont
  {M{\"a}kinen}}, \bibinfo {author} {\bibfnamefont {G.~E.}\ \bibnamefont
  {Volovik}}, \bibinfo {author} {\bibfnamefont {V.~V.}\ \bibnamefont
  {Zavjalov}},\ and\ \bibinfo {author} {\bibfnamefont {V.~B.}\ \bibnamefont
  {Eltsov}},\ }\bibfield  {title} {\bibinfo {title} {Ac josephson effect
  between two superfluid time crystals},\ }\href
  {https://doi.org/10.1038/s41563-020-0780-y} {\bibfield  {journal} {\bibinfo
  {journal} {Nature Materials}\ }\textbf {\bibinfo {volume} {20}},\ \bibinfo
  {pages} {171} (\bibinfo {year} {2021})}\BibitemShut {NoStop}%
\bibitem [{\citenamefont {Taheri}\ \emph {et~al.}(2022)\citenamefont {Taheri},
  \citenamefont {Matsko}, \citenamefont {Maleki},\ and\ \citenamefont
  {Sacha}}]{taheri22all}%
  \BibitemOpen
  \bibfield  {author} {\bibinfo {author} {\bibfnamefont {H.}~\bibnamefont
  {Taheri}}, \bibinfo {author} {\bibfnamefont {A.~B.}\ \bibnamefont {Matsko}},
  \bibinfo {author} {\bibfnamefont {L.}~\bibnamefont {Maleki}},\ and\ \bibinfo
  {author} {\bibfnamefont {K.}~\bibnamefont {Sacha}},\ }\bibfield  {title}
  {\bibinfo {title} {All-optical dissipative discrete time crystals},\ }\href
  {https://doi.org/10.1038/s41467-022-28462-x} {\bibfield  {journal} {\bibinfo
  {journal} {Nature Communications}\ }\textbf {\bibinfo {volume} {13}},\
  \bibinfo {pages} {848} (\bibinfo {year} {2022})}\BibitemShut {NoStop}%
\bibitem [{\citenamefont {Carollo}\ \emph {et~al.}(2020)\citenamefont
  {Carollo}, \citenamefont {Brandner},\ and\ \citenamefont
  {Lesanovsky}}]{carollo20nonequilibrium}%
  \BibitemOpen
  \bibfield  {author} {\bibinfo {author} {\bibfnamefont {F.}~\bibnamefont
  {Carollo}}, \bibinfo {author} {\bibfnamefont {K.}~\bibnamefont {Brandner}},\
  and\ \bibinfo {author} {\bibfnamefont {I.}~\bibnamefont {Lesanovsky}},\
  }\bibfield  {title} {\bibinfo {title} {Nonequilibrium many-body quantum
  engine driven by time-translation symmetry breaking},\ }\href
  {https://doi.org/10.1103/PhysRevLett.125.240602} {\bibfield  {journal}
  {\bibinfo  {journal} {Phys. Rev. Lett.}\ }\textbf {\bibinfo {volume} {125}},\
  \bibinfo {pages} {240602} (\bibinfo {year} {2020})}\BibitemShut {NoStop}%
\bibitem [{\citenamefont {Koch}\ \emph {et~al.}(2023)\citenamefont {Koch},
  \citenamefont {Menon}, \citenamefont {Cuestas}, \citenamefont {Barbosa},
  \citenamefont {Lutz}, \citenamefont {Fogarty}, \citenamefont {Busch},\ and\
  \citenamefont {Widera}}]{koch23a}%
  \BibitemOpen
  \bibfield  {author} {\bibinfo {author} {\bibfnamefont {J.}~\bibnamefont
  {Koch}}, \bibinfo {author} {\bibfnamefont {K.}~\bibnamefont {Menon}},
  \bibinfo {author} {\bibfnamefont {E.}~\bibnamefont {Cuestas}}, \bibinfo
  {author} {\bibfnamefont {S.}~\bibnamefont {Barbosa}}, \bibinfo {author}
  {\bibfnamefont {E.}~\bibnamefont {Lutz}}, \bibinfo {author} {\bibfnamefont
  {T.}~\bibnamefont {Fogarty}}, \bibinfo {author} {\bibfnamefont
  {T.}~\bibnamefont {Busch}},\ and\ \bibinfo {author} {\bibfnamefont
  {A.}~\bibnamefont {Widera}},\ }\bibfield  {title} {\bibinfo {title} {A
  quantum engine in the bec--bcs crossover},\ }\href
  {https://doi.org/10.1038/s41586-023-06469-8} {\bibfield  {journal} {\bibinfo
  {journal} {Nature}\ }\textbf {\bibinfo {volume} {621}},\ \bibinfo {pages}
  {723} (\bibinfo {year} {2023})}\BibitemShut {NoStop}%
\bibitem [{\citenamefont {Klatzow}\ \emph {et~al.}(2019)\citenamefont
  {Klatzow}, \citenamefont {Becker}, \citenamefont {Ledingham}, \citenamefont
  {Weinzetl}, \citenamefont {Kaczmarek}, \citenamefont {Saunders},
  \citenamefont {Nunn}, \citenamefont {Walmsley}, \citenamefont {Uzdin},\ and\
  \citenamefont {Poem}}]{klatzow19experimental}%
  \BibitemOpen
  \bibfield  {author} {\bibinfo {author} {\bibfnamefont {J.}~\bibnamefont
  {Klatzow}}, \bibinfo {author} {\bibfnamefont {J.~N.}\ \bibnamefont {Becker}},
  \bibinfo {author} {\bibfnamefont {P.~M.}\ \bibnamefont {Ledingham}}, \bibinfo
  {author} {\bibfnamefont {C.}~\bibnamefont {Weinzetl}}, \bibinfo {author}
  {\bibfnamefont {K.~T.}\ \bibnamefont {Kaczmarek}}, \bibinfo {author}
  {\bibfnamefont {D.~J.}\ \bibnamefont {Saunders}}, \bibinfo {author}
  {\bibfnamefont {J.}~\bibnamefont {Nunn}}, \bibinfo {author} {\bibfnamefont
  {I.~A.}\ \bibnamefont {Walmsley}}, \bibinfo {author} {\bibfnamefont
  {R.}~\bibnamefont {Uzdin}},\ and\ \bibinfo {author} {\bibfnamefont
  {E.}~\bibnamefont {Poem}},\ }\bibfield  {title} {\bibinfo {title}
  {Experimental demonstration of quantum effects in the operation of
  microscopic heat engines},\ }\href
  {https://doi.org/10.1103/PhysRevLett.122.110601} {\bibfield  {journal}
  {\bibinfo  {journal} {Phys. Rev. Lett.}\ }\textbf {\bibinfo {volume} {122}},\
  \bibinfo {pages} {110601} (\bibinfo {year} {2019})}\BibitemShut {NoStop}%
\bibitem [{\citenamefont {Peterson}\ \emph {et~al.}(2019)\citenamefont
  {Peterson}, \citenamefont {Batalh\~ao}, \citenamefont {Herrera},
  \citenamefont {Souza}, \citenamefont {Sarthour}, \citenamefont {Oliveira},\
  and\ \citenamefont {Serra}}]{peterson19experimental}%
  \BibitemOpen
  \bibfield  {author} {\bibinfo {author} {\bibfnamefont {J.~P.~S.}\
  \bibnamefont {Peterson}}, \bibinfo {author} {\bibfnamefont {T.~B.}\
  \bibnamefont {Batalh\~ao}}, \bibinfo {author} {\bibfnamefont
  {M.}~\bibnamefont {Herrera}}, \bibinfo {author} {\bibfnamefont {A.~M.}\
  \bibnamefont {Souza}}, \bibinfo {author} {\bibfnamefont {R.~S.}\ \bibnamefont
  {Sarthour}}, \bibinfo {author} {\bibfnamefont {I.~S.}\ \bibnamefont
  {Oliveira}},\ and\ \bibinfo {author} {\bibfnamefont {R.~M.}\ \bibnamefont
  {Serra}},\ }\bibfield  {title} {\bibinfo {title} {Experimental
  characterization of a spin quantum heat engine},\ }\href
  {https://doi.org/10.1103/PhysRevLett.123.240601} {\bibfield  {journal}
  {\bibinfo  {journal} {Phys. Rev. Lett.}\ }\textbf {\bibinfo {volume} {123}},\
  \bibinfo {pages} {240601} (\bibinfo {year} {2019})}\BibitemShut {NoStop}%
\bibitem [{\citenamefont {Niedenzu}\ and\ \citenamefont
  {Kurizki}(2018)}]{niedenzu18cooperative}%
  \BibitemOpen
  \bibfield  {author} {\bibinfo {author} {\bibfnamefont {W.}~\bibnamefont
  {Niedenzu}}\ and\ \bibinfo {author} {\bibfnamefont {G.}~\bibnamefont
  {Kurizki}},\ }\bibfield  {title} {\bibinfo {title} {Cooperative many-body
  enhancement of quantum thermal machine power},\ }\href
  {https://doi.org/10.1088/1367-2630/aaed55} {\bibfield  {journal} {\bibinfo
  {journal} {New Journal of Physics}\ }\textbf {\bibinfo {volume} {20}},\
  \bibinfo {pages} {113038} (\bibinfo {year} {2018})}\BibitemShut {NoStop}%
\bibitem [{\citenamefont {Jaseem}\ \emph {et~al.}(2023)\citenamefont {Jaseem},
  \citenamefont {Vinjanampathy},\ and\ \citenamefont
  {Mukherjee}}]{jaseem23quadratic}%
  \BibitemOpen
  \bibfield  {author} {\bibinfo {author} {\bibfnamefont {N.}~\bibnamefont
  {Jaseem}}, \bibinfo {author} {\bibfnamefont {S.}~\bibnamefont
  {Vinjanampathy}},\ and\ \bibinfo {author} {\bibfnamefont {V.}~\bibnamefont
  {Mukherjee}},\ }\bibfield  {title} {\bibinfo {title} {Quadratic enhancement
  in the reliability of collective quantum engines},\ }\href
  {https://doi.org/10.1103/PhysRevA.107.L040202} {\bibfield  {journal}
  {\bibinfo  {journal} {Phys. Rev. A}\ }\textbf {\bibinfo {volume} {107}},\
  \bibinfo {pages} {L040202} (\bibinfo {year} {2023})}\BibitemShut {NoStop}%
\bibitem [{\citenamefont {Norcia}\ \emph {et~al.}(2018)\citenamefont {Norcia},
  \citenamefont {Lewis-Swan}, \citenamefont {Cline}, \citenamefont {Zhu},
  \citenamefont {Rey},\ and\ \citenamefont {Thompson}}]{norcia18cavity}%
  \BibitemOpen
  \bibfield  {author} {\bibinfo {author} {\bibfnamefont {M.~A.}\ \bibnamefont
  {Norcia}}, \bibinfo {author} {\bibfnamefont {R.~J.}\ \bibnamefont
  {Lewis-Swan}}, \bibinfo {author} {\bibfnamefont {J.~R.~K.}\ \bibnamefont
  {Cline}}, \bibinfo {author} {\bibfnamefont {B.}~\bibnamefont {Zhu}}, \bibinfo
  {author} {\bibfnamefont {A.~M.}\ \bibnamefont {Rey}},\ and\ \bibinfo {author}
  {\bibfnamefont {J.~K.}\ \bibnamefont {Thompson}},\ }\bibfield  {title}
  {\bibinfo {title} {Cavity-mediated collective spin-exchange interactions in a
  strontium superradiant laser},\ }\href
  {https://doi.org/10.1126/science.aar3102} {\bibfield  {journal} {\bibinfo
  {journal} {Science}\ }\textbf {\bibinfo {volume} {361}},\ \bibinfo {pages}
  {259} (\bibinfo {year} {2018})},\ \Eprint
  {https://arxiv.org/abs/https://www.science.org/doi/pdf/10.1126/science.aar3102}
  {https://www.science.org/doi/pdf/10.1126/science.aar3102} \BibitemShut
  {NoStop}%
\bibitem [{\citenamefont {Campisi}\ and\ \citenamefont
  {Fazio}(2016)}]{campisi16the}%
  \BibitemOpen
  \bibfield  {author} {\bibinfo {author} {\bibfnamefont {M.}~\bibnamefont
  {Campisi}}\ and\ \bibinfo {author} {\bibfnamefont {R.}~\bibnamefont
  {Fazio}},\ }\bibfield  {title} {\bibinfo {title} {The power of a critical
  heat engine},\ }\href {https://doi.org/10.1038/ncomms11895} {\bibfield
  {journal} {\bibinfo  {journal} {Nature Communications}\ }\textbf {\bibinfo
  {volume} {7}},\ \bibinfo {pages} {11895} (\bibinfo {year}
  {2016})}\BibitemShut {NoStop}%
\bibitem [{\citenamefont {Holubec}\ and\ \citenamefont
  {Ryabov}(2017)}]{holubec17work}%
  \BibitemOpen
  \bibfield  {author} {\bibinfo {author} {\bibfnamefont {V.}~\bibnamefont
  {Holubec}}\ and\ \bibinfo {author} {\bibfnamefont {A.}~\bibnamefont
  {Ryabov}},\ }\bibfield  {title} {\bibinfo {title} {Work and power
  fluctuations in a critical heat engine},\ }\href
  {https://doi.org/10.1103/PhysRevE.96.030102} {\bibfield  {journal} {\bibinfo
  {journal} {Phys. Rev. E}\ }\textbf {\bibinfo {volume} {96}},\ \bibinfo
  {pages} {030102} (\bibinfo {year} {2017})}\BibitemShut {NoStop}%
\bibitem [{\citenamefont {Zaletel}\ \emph {et~al.}(2023)\citenamefont
  {Zaletel}, \citenamefont {Lukin}, \citenamefont {Monroe}, \citenamefont
  {Nayak}, \citenamefont {Wilczek},\ and\ \citenamefont
  {Yao}}]{zaletel23colloquium}%
  \BibitemOpen
  \bibfield  {author} {\bibinfo {author} {\bibfnamefont {M.~P.}\ \bibnamefont
  {Zaletel}}, \bibinfo {author} {\bibfnamefont {M.}~\bibnamefont {Lukin}},
  \bibinfo {author} {\bibfnamefont {C.}~\bibnamefont {Monroe}}, \bibinfo
  {author} {\bibfnamefont {C.}~\bibnamefont {Nayak}}, \bibinfo {author}
  {\bibfnamefont {F.}~\bibnamefont {Wilczek}},\ and\ \bibinfo {author}
  {\bibfnamefont {N.~Y.}\ \bibnamefont {Yao}},\ }\bibfield  {title} {\bibinfo
  {title} {Colloquium: Quantum and classical discrete time crystals},\ }\href
  {https://doi.org/10.1103/RevModPhys.95.031001} {\bibfield  {journal}
  {\bibinfo  {journal} {Rev. Mod. Phys.}\ }\textbf {\bibinfo {volume} {95}},\
  \bibinfo {pages} {031001} (\bibinfo {year} {2023})}\BibitemShut {NoStop}%
\bibitem [{\citenamefont {B.S}\ \emph {et~al.}(2022)\citenamefont {B.S},
  \citenamefont {Mukherjee},\ and\ \citenamefont {Divakaran}}]{revathy22bath}%
  \BibitemOpen
  \bibfield  {author} {\bibinfo {author} {\bibfnamefont {R.}~\bibnamefont
  {B.S}}, \bibinfo {author} {\bibfnamefont {V.}~\bibnamefont {Mukherjee}},\
  and\ \bibinfo {author} {\bibfnamefont {U.}~\bibnamefont {Divakaran}},\
  }\bibfield  {title} {\bibinfo {title} {Bath engineering enhanced quantum
  critical engines},\ }\bibfield  {journal} {\bibinfo  {journal} {Entropy}\
  }\textbf {\bibinfo {volume} {24}},\ \href {https://doi.org/10.3390/e24101458}
  {10.3390/e24101458} (\bibinfo {year} {2022})\BibitemShut {NoStop}%
\bibitem [{\citenamefont {Yunt}\ \emph {et~al.}(2020)\citenamefont {Yunt},
  \citenamefont {Fadaie}, \citenamefont {M\"ustecapl\ifmmode \imath \else \i
  \fi{}o\ifmmode~\breve{g}\else \u{g}\fi{}lu},\ and\ \citenamefont
  {Smith}}]{yunt20internal}%
  \BibitemOpen
  \bibfield  {author} {\bibinfo {author} {\bibfnamefont {E.}~\bibnamefont
  {Yunt}}, \bibinfo {author} {\bibfnamefont {M.}~\bibnamefont {Fadaie}},
  \bibinfo {author} {\bibfnamefont {O.~E.}\ \bibnamefont {M\"ustecapl\ifmmode
  \imath \else \i \fi{}o\ifmmode~\breve{g}\else \u{g}\fi{}lu}},\ and\ \bibinfo
  {author} {\bibfnamefont {C.~M.}\ \bibnamefont {Smith}},\ }\bibfield  {title}
  {\bibinfo {title} {Internal geometric friction in a kitaev-chain heat
  engine},\ }\href {https://doi.org/10.1103/PhysRevB.102.155423} {\bibfield
  {journal} {\bibinfo  {journal} {Phys. Rev. B}\ }\textbf {\bibinfo {volume}
  {102}},\ \bibinfo {pages} {155423} (\bibinfo {year} {2020})}\BibitemShut
  {NoStop}%
\bibitem [{\citenamefont {Campbell}\ and\ \citenamefont
  {Deffner}(2017)}]{campbell17trade}%
  \BibitemOpen
  \bibfield  {author} {\bibinfo {author} {\bibfnamefont {S.}~\bibnamefont
  {Campbell}}\ and\ \bibinfo {author} {\bibfnamefont {S.}~\bibnamefont
  {Deffner}},\ }\bibfield  {title} {\bibinfo {title} {Trade-off between speed
  and cost in shortcuts to adiabaticity},\ }\href
  {https://doi.org/10.1103/PhysRevLett.118.100601} {\bibfield  {journal}
  {\bibinfo  {journal} {Phys. Rev. Lett.}\ }\textbf {\bibinfo {volume} {118}},\
  \bibinfo {pages} {100601} (\bibinfo {year} {2017})}\BibitemShut {NoStop}%
\bibitem [{\citenamefont {Bai}\ and\ \citenamefont {An}(2020)}]{bai20floquet}%
  \BibitemOpen
  \bibfield  {author} {\bibinfo {author} {\bibfnamefont {S.-Y.}\ \bibnamefont
  {Bai}}\ and\ \bibinfo {author} {\bibfnamefont {J.-H.}\ \bibnamefont {An}},\
  }\bibfield  {title} {\bibinfo {title} {Floquet engineering to reactivate a
  dissipative quantum battery},\ }\href
  {https://doi.org/10.1103/PhysRevA.102.060201} {\bibfield  {journal} {\bibinfo
   {journal} {Phys. Rev. A}\ }\textbf {\bibinfo {volume} {102}},\ \bibinfo
  {pages} {060201(R)} (\bibinfo {year} {2020})}\BibitemShut {NoStop}%
\end{thebibliography}
\end{document}